\documentstyle[11pt,epsfig]{article}
 
\def\fig#1{Fig.~{\ref{#1}}}
\def\eqn#1{Eq.~(\ref{#1})}

\newskip\humongous \humongous=0pt plus 1000pt minus 1000pt
\def\caja{\mathsurround=0pt}
\def\eqalign#1{\,\vcenter{\openup1\jot \caja
        \ialign{\strut \hfil$\displaystyle{##}$&$
        \displaystyle{{}##}$\hfil\crcr#1\crcr}}\,}
\newif\ifdtup

\newcounter{eqnumber}
\renewcommand{\theeqnumber}{\arabic{eqnumber}}
\def\equn{
\refstepcounter{eqnumber}
\eqno({\rm \theeqnumber})
}

\def\eqn#1{eq.~(\ref{#1})}

\def\fig#1{fig.~{\ref{#1}}}

\newbox\charbox
\newbox\slabox
\def\s#1{{      
        \setbox\charbox=\hbox{$#1$}
        \setbox\slabox=\hbox{$/$}
        \dimen\charbox=\ht\slabox
        \advance\dimen\charbox by -\dp\slabox
        \advance\dimen\charbox by -\ht\charbox
        \advance\dimen\charbox by \dp\charbox
        \divide\dimen\charbox by 2
        \raise-\dimen\charbox\hbox to \wd\charbox{\hss/\hss}
        \llap{$#1$}
}}

\def\spa#1.#2{\left\langle#1\,#2\right\rangle}
\def\spb#1.#2{\left[#1\,#2\right]}
\def\lor#1.#2{\left(#1\,#2\right)}
\def\sand#1.#2.#3{%
  \left\langle\smash{#1}{\vphantom1}\right|{#2}%
  \left|\smash{#3}{\vphantom1}\right\rangle}
\def\sandp#1.#2.#3{%
  \left\langle\smash{#1}{\vphantom1}^{-}\right|{#2}%
  \left|\smash{#3}{\vphantom1}^{+}\right\rangle}
\def\sandpp#1.#2.#3{%
  \left\langle\smash{#1}{\vphantom1}^{+}\right|{#2}%
  \left|\smash{#3}{\vphantom1}^{+}\right\rangle}
\def\sandmm#1.#2.#3{%
  \left\langle\smash{#1}{\vphantom1}^{-}\right|{#2}%
  \left|\smash{#3}{\vphantom1}^{-}\right\rangle}
\def\sandpm#1.#2.#3{%
  \left\langle\smash{#1}{\vphantom1}^{+}\right|{#2}%
  \left|\smash{#3}{\vphantom1}^{-}\right\rangle}
\def\sandmp#1.#2.#3{%
  \left\langle\smash{#1}{\vphantom1}^{-}\right|{#2}%
  \left|\smash{#3}{\vphantom1}^{+}\right\rangle}
\catcode`@=11  

\def\Li{\, {\rm Li}}

\def\eps{\epsilon}

\def\pol{\eps}
\def\sandmy#1.#2{\left\langle#1 | #2 \right\rangle}
\def\N{{\tilde{\chi}}}
\def\ch{{\tilde{\chi}^\pm}}

\def\Re{\mathop{\rm Re}}

\begin{document}

\begin{titlepage}

\begin{flushright}
hep-ph/9706538 \hfill UCLA/97/TEP/11\\
\hfill MPI-PhT/97-36\\
June, 1997
\end{flushright}

\vskip 1.cm

\begin{center}
{\Large\bf Neutralino Annihilation into Two Photons}
\vskip 2.cm

{Zvi Bern\footnote{bern@physics.ucla.edu}}
\vskip 0.2cm

{\it  Department of Physics\\
University of California at Los Angeles\\
Los Angeles,  CA 90095-1547}
\vskip .5cm

{Paolo Gondolo\footnote{gondolo@mppmu.mpg.de}}
\vskip 0.2cm

{\it Max Planck Institut f{\"u}r Physik\\
 F{\"o}hringer Ring 6, 80805 M{\"u}nchen,
  Germany }
\vskip .3cm

and

\vskip .3 cm 
{Maxim Perelstein\footnote{maxim@physics.ucla.edu}}

\vskip 0.2cm

{\it  Department of Physics\\
University of California at Los Angeles\\
Los Angeles,  CA 90095-1547}
\vskip 3cm
\end{center}

\begin{abstract}
  We compute the annihilation cross-section of two neutralinos at rest
  into two photons, which is of importance for the indirect detection
  of neutralino dark matter in the galactic halo through a
  quasi-monochromatic gamma-ray line.  We include all diagrams to
  one-loop level in the minimal supersymmetric extension of the
  standard model. We use the helicity formalism, the background-field
  gauge, and an efficient loop-integral reduction method. We confirm
  the result recently obtained by Bergstr{\"o}m and Ullio in a different
  gauge, which disagrees with other published calculations.
\end{abstract}

\vfill
\end{titlepage}

\baselineskip 16pt
\section{Introduction}

The composition of dark matter is one of the major issues in
astrophysics. A popular candidate for non-baryonic dark matter is a
neutral stable Majorana fermion, the lightest neutralino, appearing in
a large class of supersymmetric extensions of the Standard Model.  In
a wide range of supersymmetric parameter space, relic neutralinos from
the Big Bang are in principle abundant enough to account for the dark
matter in our galactic halo. (See ref.~\cite{EG} for a thorough
analysis and ref.~\cite{JKG} for a comprehensive review.)

Several methods have been proposed to detect galactic neutralinos:
elastic scattering in a low-background terrestrial detector, energetic
neutrino fluxes from neutralino annihilations in the core of the Sun
or of the Earth, gamma-rays and cosmic ray antiprotons or positrons
from neutralino annihilation in the halo. (For a review see
ref.~\cite{JKG}.)

In particular, the annihilation into two photons gives a distinct
signature against backgrounds from known astrophysical sources -- a
narrow line in the gamma-ray spectrum at the energy equal to the
neutralino mass \cite{Bergstrom}.  Observation of such a line would
provide evidence for the existence of supersymmetric dark matter,
while non-observation can be used to put constraints on models of the
galactic dark matter.  A number of new experiments with reasonable
energy resolutions are being planned \cite{Experiments} which would be
sensitive to photon production in the galactic halo.

There exist several previous calculations of the neutralino
annihilation into two photons, all but one incomplete.  The first
calculation was carried out by Bergstr{\"o}m and Snellman
\cite{Bergstrom} for neutralinos which are pure photinos. Rudaz
\cite{History1} corrected their results and considered also pure
higgsinos.  Giudice and Griest \cite{History2} removed the restriction
on the neutralino state mixture, and confirmed Rudaz's work.  In all
these early calculations the sfermion and the $W$-boson masses were
assumed to be much larger than any other mass. This assumption was
relaxed by Bergstr{\"o}m \cite{History3}, who computed the
fermion-sfermion loop contributions for arbitrary squark masses, but
only for a pure photino.  Bergstr{\"o}m and Kaplan \cite{History4}
evaluated the contribution from virtual $W$'s in a leading-logarithm
approximation. Jungman and Kamionkowski \cite{JK} improved on the
leading-logarithm approximation, but did not include all contributing
diagrams (our $A^b_{W \psi}$ below was missing). Only very recently,
Bergstr{\"o}m and Ullio \cite{BU} (hereafter BU) presented a full
one-loop calculation. Their results, however, differ from some
previous partial calculations when corresponding expressions are
compared.

We provide an independent complete one-loop calculation of the
neutralino annihilation into two photons. We confirm the BU result
using a different calculational procedure.  
To perform the calculation we take advantage of some recent
improvements in the calculational techniques for one-loop amplitudes
\cite{Review}.  In particular, we use a helicity basis
\cite{SpinorHelicity,XZC} for the photons, background field
Feynman-'t~Hooft gauge~\cite{Background,WeakBackground} and an
efficient integral reduction method \cite{IntegralEvaluate}.

As in previous calculations, we focus only on the first term in the
relativistic expansion of $\sigma v$, where $\sigma$ is the
annihilation cross-section and $v$ the relative neutralino-neutralino
velocity. This kinematic limit is appropriate for neutralino
annihilation in the galactic halo, where the neutralino velocities are
of the order of the galactic rotation speed, ${v /c} \simeq
10^{-3}$. This kinematic configuration considerably simplifies the
form of the amplitudes. Our evaluation methods are however general and
do not depend on the special kinematics.


\section{Evaluation of the Amplitudes}

\begin{figure}[ht]
\begin{center}
\vskip -.7 cm
\epsfig{file=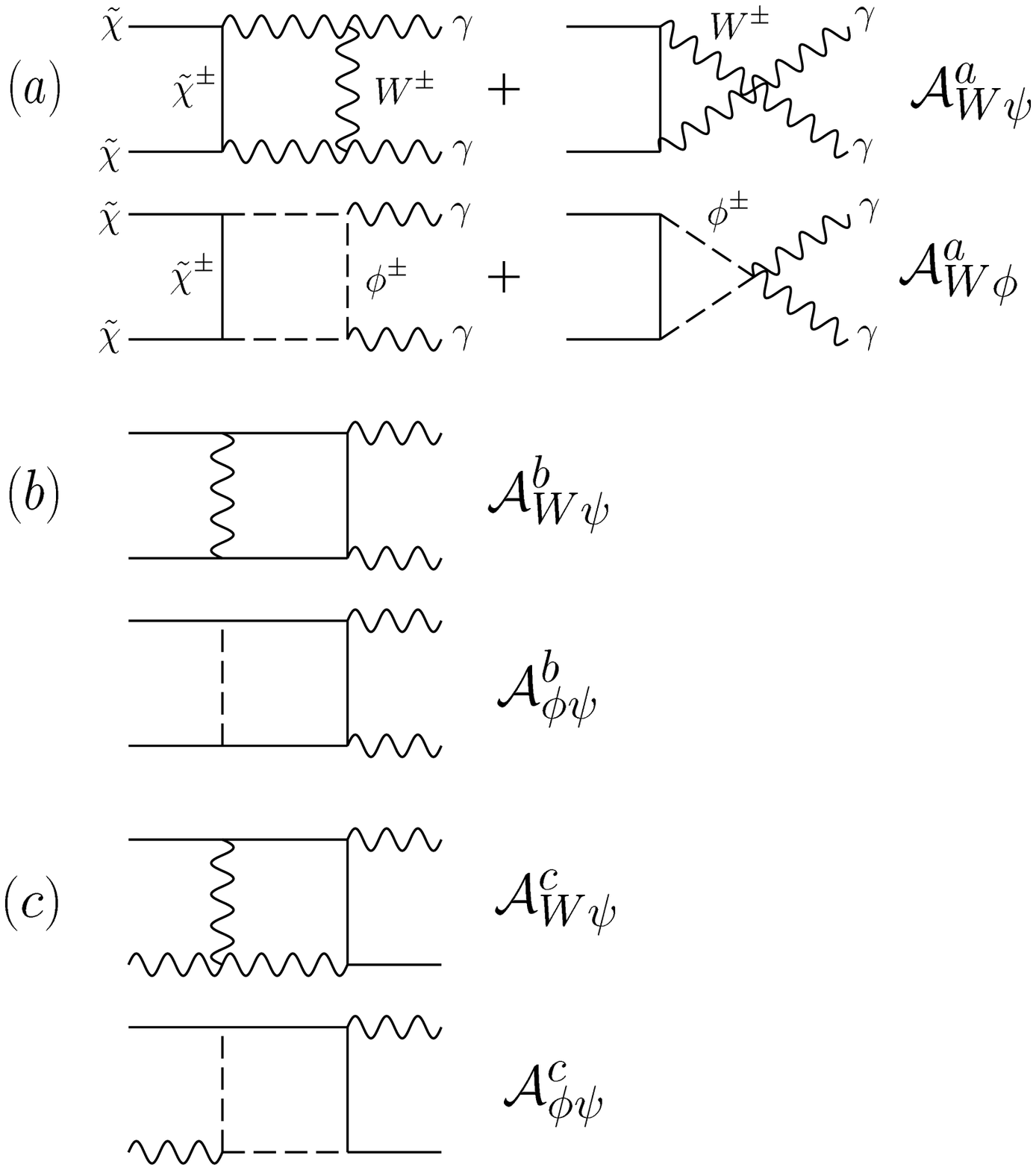,clip=,width=3.2in}
\epsfig{file=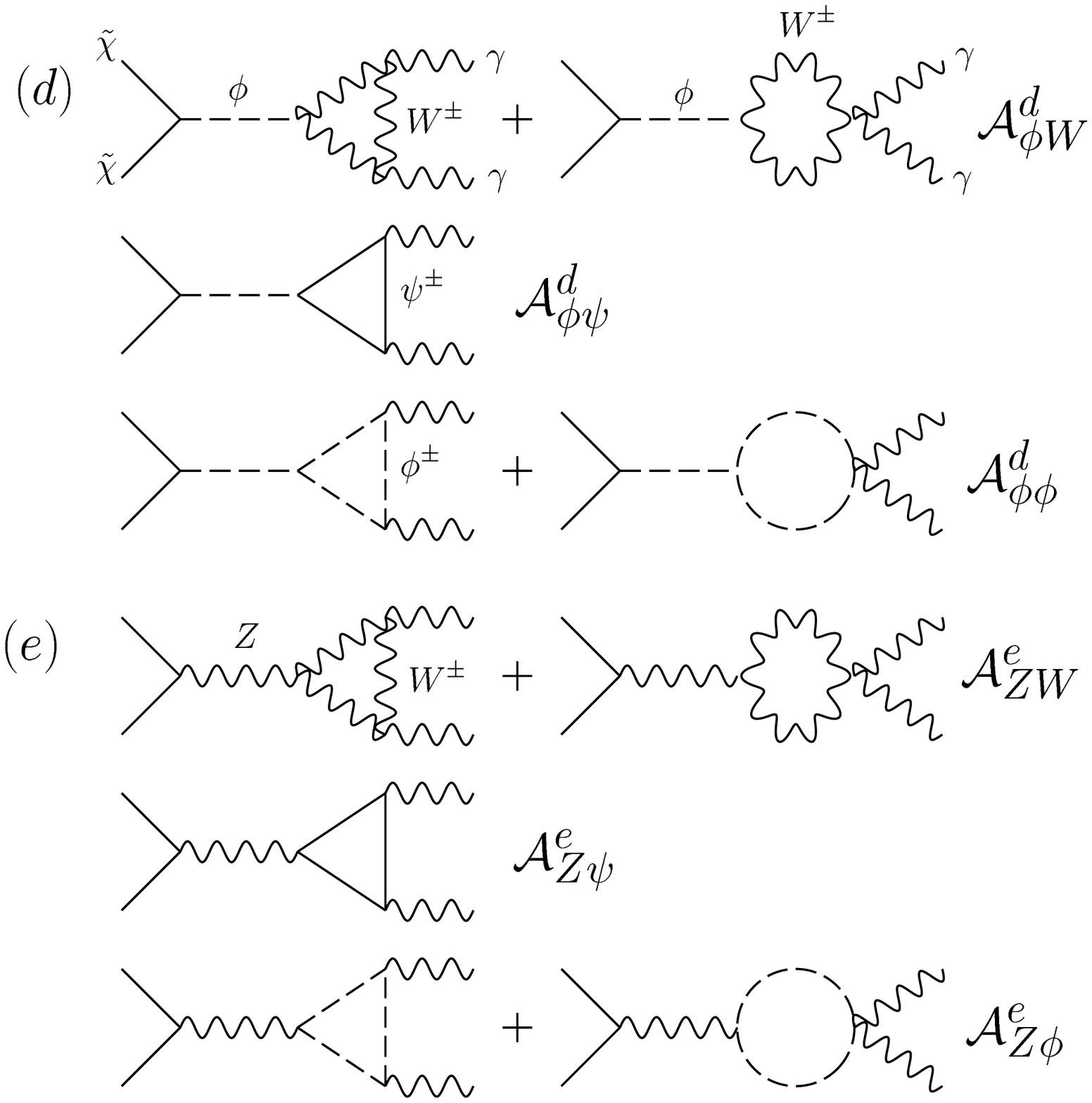,clip=,width=3.2in}
\end{center}
\vskip -.75 cm
\caption[]{
\label{Diagrams}
\small
Feynman diagrams for neutralino annihilation into two photons.}
\end{figure}

In this section we outline the computation of the amplitudes for the
annihilation of two neutral massive Majorana fermions at rest into two
photons via vector boson, scalar and fermion loops.  The Feynman
diagrams for the process under consideration are given in
\fig{Diagrams}.  Since photons do not change the identity of the
particle they couple to, the masses in the fermion and scalar lines
within the loops are uniform.

\subsection{Feynman Rules and Diagrams}

\begin{figure}[ht]
\begin{center}
\vskip -.7 cm
\epsfig{file=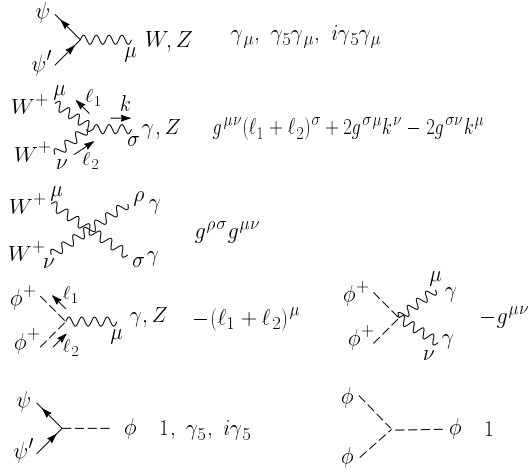,clip=,width=2.8in}
\end{center}
\vskip -.75 cm
\caption[]{
\label{Rules}
\small Feynman rules in the background field gauge; overall factors of
$i$ and coupling constants have been removed.}
\end{figure}

We use background field Feynman--'t Hooft gauge
\cite{Background,WeakBackground} because of its technical
advantages.  In this gauge the gauge fixing part of the Lagrangian is
$$
\eqalign{
{\cal L}_{gf} 
& = -{1 \over 2} \left( \partial_{\mu} W^{i \mu} + g \epsilon^{ijk} \, 
\tilde{W}_{j \mu} W_k^{\mu} + {ig \over 2} \sum (\phi^{\prime\dagger}_k T^i 
\phi_{0k} - \phi_{0k}^{\dagger} T^i \phi^{\prime}_{k}) \right)^2 \cr
&\hskip 2 cm 
-{1 \over 2} \left( \partial_\mu B^{\mu}+{ig^\prime \over 2} 
\sum (
\phi^{\prime\dagger}_k \phi_{0k} - \phi_{0k}^{\dagger} \phi_k^\prime)
\right)^2 \,, \cr
}
\equn
$$
where $\tilde{W}$ and $\tilde{B}$ are respectively the $SU(2)$ and
$U(1)$ external background fields, $W$ and $B$ are the corresponding
quantum fields appearing in the loops, and $\phi_{0k}$ is the VEV of
the $k$-th Higgs field of the model: $\phi_k =
\phi_{0k}+\phi_k^{\prime}$.  The Fadeev-Popov ghosts are rather simple
and at one-loop have the same coupling as scalars; only the overall
sign differs because of Fermi statistics.  The Feynman rules for
performing this calculation may be found in
ref.~\cite{WeakBackground} and are summarized in \fig{Rules}.  In this
figure $\gamma$ and $Z$ are background fields.

An important technical advantage of this gauge is that only one of the
three terms in the $W W \gamma$ vertex contains loop momentum.  This
decreases the number of the integrals with high powers of loop
momentum in the numerator that have to be evaluated. Typically, such
integrals are the most difficult to compute.  Furthermore, the gauge
fixing terms have been chosen so that they cancel the $W H \gamma$
coupling from the Lagrangian, which reduces the number of diagrams to
be evaluated. (This is similar to non-linear $R_{\xi}$ gauges
\cite{Nonlinear,History4,JK}.) This cancellation works for any number
of Higgs fields assuming that they are doublets of the Standard Model
$SU(2)$.  For the massive four-point amplitudes computed in this
paper, the background field method captures the main technical
advantages of using the gauge invariant cutting techniques reviewed in
ref.~\cite{Review}.

\subsection{Helicity and Kinematics}
\label{sect:kinematics}

We use a helicity representation for the final-state photons, since
this generally leads to compact expressions for amplitudes.  In the
spinor helicity formalism~\cite{SpinorHelicity,XZC} the photon
polarization vectors are expressed in terms of massless Weyl spinors
$|\,k^\pm \rangle = {1\over 2} (1\pm \gamma_5) |\, k\, \rangle$,
$$
\pol^{+}_\mu (k;q) = 
     {\sandmm{q}.{\gamma_\mu}.k
      \over \sqrt2 \spa{q}.k}\, ,\hskip 1cm
\pol^{-}_\mu (k;q) = 
     {\sandpp{q}.{\gamma_\mu}.k
      \over \sqrt{2} \spb{k}.q} \, ,
\equn\label{HelicityDef}
$$  
where $k$ is the photon momentum, $q$ is an arbitrary null
`reference momentum' which drops out of final gauge-invariant
amplitudes and
$$
\spa{i}.j \equiv  \langle k_i^{-} \vert k_j^{+} \rangle\,, 
\hskip 2.0 cm 
\spb{i}.j \equiv \langle k_i^{+} \vert k_j^{-} \rangle \,,
\equn
$$
which satisfy the normalization condition $\spa{i}.{j} \spb{j}.{i} = 2
k_i \cdot k_j$.  The plus and minus labels on the polarization vectors
refer to the outgoing photon helicities.  We denote the massive
Majorana spinors of the initial-state neutralinos by $|\, 1 \rangle$
and $|\, 2 \rangle$.  (For the massive spinors we choose the 
normalizations of ref.\cite{SpinorNormalizations}; our
massless spinors follow the conventions of ref.~\cite{XZC}.)

Since the two annihilating particles are identical Majorana fermions,
the initial-state wave-function must be antisymmetric. Therefore the
initial spin state must be the singlet with angular momentum zero;
since it is conserved, the final-state photons must also have
vanishing angular momentum and be of the same helicity.  This means
that there are only two helicity cases that are required, namely where
both photons are of positive or both are of negative helicity.  These
two helicity configurations are, however, simply related as can easily
be seen by making the reference momentum choice
$$
q_3 = k_4\, , \hskip 2 cm q_4 = k_3\,,
\equn\label{ReferenceMomentum}
$$
where $q_3$ and $q_4$ are the reference momenta of the two photons.
With this choice of reference momenta
$$
\pol_4^{-\mu}={\spa3.4 \over \spb3.4}\; \pol_3^{+\mu}, 
\hskip 1.5 cm \pol_3^{-\mu}={\spa3.4 \over \spb3.4}\; \pol_4^{+\mu}.
\equn
$$ 
Since the amplitudes are symmetric under the interchange of 
photons $3 \leftrightarrow 4$, we have
$$
{\cal A}(1_\N,2_\N, 3_\gamma^-, 4_\gamma^-)
 = {{\spa3.4}^2 \over {\spb3.4}^2} \,
{\cal A}(1_\N, 2_\N, 3_\gamma^+, 4_\gamma^+)\,,
\equn
$$
where legs 1 and 2 are the initial state neutralinos and the $\pm$
labels on legs $3$ and $4$ refer to the photon helicities.  Thus, the
amplitudes for either helicity choice are identical up to an overall
phase factor and we need only compute the annihilation into two
positive helicity photons to obtain the entire result.

A number of important simplifications occur because of the special 
kinematics. For annihilation at rest, we take the
kinematic configuration to be of the form
$$
\eqalign{
& k_1 = (m_\N, 0, 0, 0)\,, \hskip 2.65 cm k_2  = (m_\N, 0, 0, 0)\,, \cr
& k_3 = (-m_\N, 0, 0, m_\N)\,, \hskip 2 cm k_4 = (-m_\N, 0, 0, -m_\N)\,, \cr}
\equn\label{SpecialKinematics}
$$
where $m_\N$ is the neutralino mass, $k_1$ and $k_2$ are the
neutralino four momenta and $k_3$ and $k_4$ the photon momenta. The
Mandelstam variables are then given by
$$
s \equiv (k_1+k_2)^2 =  4 \, m_\N^2\,, \hskip 1 cm 
t \equiv (k_1+k_4)^2 =  - m_\N^2\,, \hskip 1 cm 
u \equiv (k_1+k_3)^2=  - m_\N^2 \,.
\equn
$$

The four-point kinematics allows us to reduce any spinor structure
that appears in the diagrams (before we symmetrize in the photons) to
a linear combination of $\sandmy1.2$, $\sand1.{\gamma_5}.2$,
$\sand1.{\s k_4}.2$, and $\sand1.{\gamma_5 \s k_4}.2$.  The
permutation $3 \leftrightarrow 4$ then corresponds to $k_{3}
\leftrightarrow k_{4}$, $t \leftrightarrow u$. (Note that the
reference momentum choice (\ref{ReferenceMomentum}) respects the
exchange symmetry.)  But in our kinematics $t=u$, so we only need to
symmetrize the spinors. Using $\sand1.{\s k_3+ \s k_4}.2 = 0$,
$\sand1.{\gamma_5 (\s k_3+ \s k_4)}.2=2m_\N \sand1.{\gamma_5}.2$, we
see that the answer after symmetrizing in the photons contains  only
$\sandmy1.2$ and $\sand1.{\gamma_5}.2$. At the kinematic point
(\ref{SpecialKinematics}) we have 
$\sandmy1.2=0$, so all amplitudes are proportional to
$\sand1.{\gamma_5}.2$.

Another simplification due to this kinematics is that we need only
consider pseudo-scalar Higgs particles and $Z$ bosons in the
s-channel, the other neutralino currents vanish.  This means that in
fact we do not need to calculate diagrams ${\cal A}_{\phi W}^d$, 
and in the diagrams ${\cal A}_{\phi \psi}^d$ we
need to consider only pseudo-scalar intermediate Higgs. Moreover,
diagrams ${\cal A}_{\phi \phi}^d$ do not contribute because CP
conservation (which we impose) forbids the $H^+H^-H^0_3$ and
$H^+H^-G^0$ couplings.

\subsection{Integral Reductions}

In evaluating the diagrams in \fig{Diagrams}, one encounters loop integrals
of the form, 
$$
I_n[P(\ell^\mu)]
 \equiv -i(4\pi)^{2-\eps} \int {d^{4-2\eps}\ell\over (2\pi)^{4-2\eps}} \;
{P(\ell^\mu)\over L_1^2 \, L_2^2 \, ... \, L_n^2 } \,,
\equn\label{GeneralMPoint}
$$
where
$$
L_i^2 = \ell_i^2-m_i^2\,.
\equn
$$
Here the $\ell_i$ and $m_i$ are the momentum and the mass flowing through the
$i$th loop propagator,  and $P(\ell^\mu)$ is a polynomial in the loop 
momenta.  For the box (four-point) integrals, 
$$
\ell_1 = \ell\,, \hskip 1.5 cm  \ell_2 = \ell - k_1\,, \hskip 1.5 cm 
\ell_3 = \ell - k_1 - k_2\,, \hskip 1.5 cm \ell_4 = \ell + k_4 \,.
\equn
$$
(Our sign conventions for the integrals $I_n$ are adjusted to agree
with Oldenborgh's FF program, and differ from those of
refs.~\cite{IntegralsShort,IntegralsLong,Review} for the three-point
integrals.)  Although, the amplitudes are ultraviolet finite we
include dimensional regularization because a few of the integrals
encountered in the calculation are divergent.  (We comment
that we use the `four-dimensional helicity scheme' \cite{FDH} to
maintain compatibility with the helicity formalism; however, since the
amplitudes are ultraviolet finite the specific scheme choice is 
unimportant.)

We now outline the procedure used to reduce the tensor integrals
(i.e. those with powers of loop momentum in the numerator of
eq.~({\ref{GeneralMPoint})) to linear combinations of scalar integral
functions used to express the amplitudes.  This procedure has been
used previously in refs.~\cite{IntegralEvaluate,Massive,Review}.  The
basic technique is to extract as many inverse propagators as possible
from the spinor inner products appearing in the numerators of the
integrands.  In this way we can simplify the tensor box integrals with
two or three powers of loop momenta, which are by far the most
complicated of the integrals occurring in the calculation.

As an example, consider the box diagram in \fig{Diagrams}{(a)} with a
scalar $\phi$ and a charged fermion $\psi$ in the loop, given by
$$
 \int {d^{4} \ell \over (2\pi)^4 } 
  {d^{-2\eps} \mu \over (2\pi)^{-2\eps}}\, 
{\sand1.{\s\ell_2 + m_2 + \s\mu}.2 \;
   \pol_3^+\! \cdot (\ell_4 + \ell_3) \; \pol_4^+ \!\cdot 
(\ell_1 + \ell_4) \over
(\ell_1^2 - \mu^2 - m_\phi^2) (\ell_2^2 -\mu^2 - m_\psi^2) 
(\ell_3^2 - \mu^2 - m_\phi^2) (\ell_4^2 -\mu^2 - m_\phi^2)} \,.
\equn\label{SampleIntegral}
$$
As usual when using four-dimensional helicities
\cite{Mahlon,Massive,Review}, we have broken up the loop integral into
a four-dimensional part, $\ell$, and a $(-2\eps)$-dimensional part,
$\mu$. See, for example, appendices A.2 and C of ref.~\cite{Massive}
for further details.

Using the polarization vectors (\ref{HelicityDef}), with the choice of
reference momentum (\ref{ReferenceMomentum}) the dot products of the
polarization vectors and loop momenta may be re-expressed as
$$
\pol_3^+ \!\cdot (\ell_4 + \ell_3) \;
\pol_4^+ \!\cdot (\ell_1 + \ell_4)  = 
-2 \, {\sandpp3.{\ell_4}.4 \sandpp4.{\ell_4}.3 \over \spa3.4^2} \,.
\equn
$$
We may then extract inverse propagators from the spinor products using, 
$$
\eqalign{
\sandpp3.{\s\ell_4}.4&  \sandpp4.{\s\ell_4}.3\cr
 & = \sandpp3.{\s\ell_4 \s k_4 \s\ell_4}.3 \cr
& =
2 \,k_4 \cdot \ell_4 \sandpp3.{\s\ell_4}.3 - \ell_4^2 \sandpp3.{\s k_4}.3 \cr
& = [(\ell_4^2 - m_\phi^2 -\mu^2) - (\ell_1^2 - m_\phi^2 - \mu^2)] 
           2 \, \ell_4 \cdot k_3  
    - [(\ell_4^2 - m_\phi^2 - \mu^2) + m_\phi^2 + \mu^2]
            s_{34}\,, \cr}
\equn\label{InversePropagators}
$$
where the combinations appearing in the parentheses are inverses of
propagators appearing in the loop integral (\ref{SampleIntegral}).  By
canceling propagators we can reduce the tensor box integral
(\ref{SampleIntegral}) to a linear combination of box integrals with
at most one power of loop momentum and triangle integrals with at most
two powers of loop momentum.  In some cases we can reduce the tensor
triangle integrals by again extracting inverse propagators; however,
this strategy fails for some terms because one cannot form an inverse
propagator appropriate for the triangle integral.  For example,
in eq.~(\ref{InversePropagators}), the term 
$$
(\ell_4^2 - m_\phi^2 - \mu^2) 2 \, \ell_4 \cdot k_3 = 
(\ell_4^2 - m_\phi^2 - \mu^2) \, [(\ell_3^2 - m_\phi^2 - \mu^2) - 
(\ell_4^2 - m_\phi^2 - \mu^2)]\,,
\equn
$$
contains two factors of $(\ell_4^2 - m_\phi^2 - \mu^2)$ which cannot
be completely canceled against a single propagator.  For such terms
one can use, for example, the integration formulas of
refs.~\cite{PV,IntegralsShort,IntegralsLong} to directly evaluate the
tensor integrals in terms of scalar integrals; the expressions are
relatively simple since these are triangle integrals.

The integrals with a single power of $\sl \mu$ in the numerator
vanish, while the ones containing a $\mu^2$ are proportional to
$\eps$ times a higher-dimensional integral \cite{Massive,Review},
$$
\int {d^{4} \ell \over (2\pi)^4 } 
  {d^{-2\eps} \mu \over (2\pi)^{-2\eps}}\; 
   \mu^2 \, f(\mu^2, \ell^\nu)
= -4 \pi \, \eps \int {d^{6} \ell \over (2\pi)^6 } 
  {d^{-2\eps} \mu \over (2\pi)^{-2\eps}}\;   f(\mu^2, \ell^\nu)\,.
\equn 
$$
The higher dimensional integral may be re-expressed in terms of
$D=4-2\eps$ integrals using the integral relations appearing in
refs.~\cite{IntegralsShort,IntegralsLong}.  Observe that because of
the explicit $\eps$, we need only evaluate the parts of the higher
dimensional integrals that are singular in $\eps$.  In some cases,
such as for box integrals containing a single power of $\mu^2$, the
integrals are finite and may therefore be dropped immediately since
they are suppressed by an overall power of $\eps$.  
It turns out that for the amplitudes presented in this paper,
all such higher dimension integrals cancel.

The above procedure allows us to reduce all four-point diagrams in
\fig{Diagrams} to a linear combination of scalar box, triangle and
bubble integrals.  

\subsection{The Cross-Section}

We calculate the cross-section in the framework of the minimal supersymmetric
extension of the Standard Model. (For a comprehensive discussion, see
ref.~\cite{Haber}.) We use the following notation for the MSSM interaction
terms that contribute to the diagrams in \fig{Diagrams}:
$$
\eqalign{
{\cal L}_{\rm int} \, & =\, 
W_\mu \bar{\chi} \gamma^\mu (g^V_{W\chi\psi} + g^A_{W\chi\psi}
\gamma_5) \psi +
\phi^* \bar{\chi} (g^S_{\phi\chi\psi} +
g^P_{\phi\chi\psi} \gamma_5) \psi + {\rm h.c.} 
\cr
&\, + g^A_{Z\psi\psi'} Z_\mu \bar{\psi} \gamma^\mu \gamma_5 \psi'
+ g^P_{\phi\psi\psi'} \phi^* \bar{\psi} \gamma_5 \psi' 
+ {\rm h.c.} \cr
&\, + {\textstyle{1\over2}}
 g^A_{Z\chi\chi'} Z_\mu \bar{\chi}  \gamma^\mu \gamma_5 \chi' +
{\textstyle{1\over2}} 
g^P_{\phi\chi\chi'} \phi^* \bar{\chi} \gamma^\mu \gamma_5 \chi 
+ {\rm h.c.} 
}
\equn
$$ 
Here $\chi,\chi'$ are Majorana fermions, $\psi,\psi'$ are Dirac fermions,
and $\phi$ can be a neutral ($\phi^*=\phi$) or a charged scalar. Explicit
expressions for the coupling constants $g_{ijk}$ can be obtained in
refs.~\cite{Haber,vertices}. 

We write the amplitude as
$$ 
{\cal A}(1_\N, 2_\N, 3_\gamma^+, 4_\gamma^+) = 
{ 4 i\, \alpha \, m_\N^3 \sand1.{\gamma_5}.2 \over \pi \, {\spa3.4}^2} \,
\tilde{\cal A} \,,
\equn
$$ 
where $\alpha$ is the fine structure constant. With this notation, 
$\tilde{\cal A}$ matches the corresponding expression in
refs.~\cite{JK,BU}.

The cross-section may be directly obtained from the amplitudes listed
in eq.~(\ref{Result}) below using the formula
$$
\sigma v = {\alpha^2 m_\N^2 \over 16 \pi^3}  | \tilde{\cal A}|^2 \, .
\equn\label{CrossSection}
$$
We have averaged over the four possible initial states,
summed over the two possible photon helicity configurations
and inserted the symmetry factor of ${1\over 2}$ for identical final
state photons. 
In evaluating the squared matrix elements we have used
$$
\left|\spa3.4\right| = 2 m_\N \, , \hskip 2 cm 
\left|\sand1.{\gamma_5}.2\right| = 2\sqrt{2} m_\N\,,
\equn
$$
valid for the kinematics (\ref{SpecialKinematics}).

We find the following non-vanishing contributions to the amplitude:
$$ 
\tilde{\cal A} = \sum_{\psi=\ch_1,\ch_2} \tilde{\cal A}^{a+b+c}_{W\psi} 
+  \sum_{\matrix{\scriptstyle \phi=H^\pm,G^\pm \cr 
                 \scriptstyle \psi=\ch_1,\ch_2}}
\tilde{\cal A}^{a+b+c}_{\phi\psi} +
  \sum_{\matrix{\scriptstyle \phi=\tilde{f} \cr 
                 \scriptstyle \psi=f}}
\tilde{\cal A}^{a+b+c}_{\phi\psi} +
 \sum_{\matrix{\scriptstyle \phi=H^0_3,G^0 \cr 
               \scriptstyle \psi=f,\ch_1,\ch_2}}
\tilde{\cal A}^{d}_{\phi\psi} + 
\sum_{\psi=f,\ch_1,\ch_2} \tilde{\cal A}^{e}_{Z\psi} 
\equn
$$ where the sums are over all the Standard Model fermions $f$ (quarks and
leptons), the corresponding sfermions $\tilde{f}$, the charginos $\ch_1,\ch_2$,
and the unphysical Higgs bosons $G^0, G^\pm$ with masses $m_{G^0}=m_Z$ and
$m_{G^\pm}=m_W$, which appear in 't Hooft-Feynman
gauge.  The individual amplitudes are
$$
\eqalign{
& \tilde{\cal A}^{a+b+c}_{W\psi}  = 
- 2\, S_{W\N\psi} 
\left[ 2 \, I_3^{[1]}(m_W) +
2 (m_{\psi}^2+m_W^2-m_\N^2)\, I_4^{[a]}
 +2 m_{\psi}^2 \, I_4^{[b]}
+ 3 m_{\psi}^2 \, I_4^{[c]} + I_3^{[2]}(m_W, m_\psi) \right]  
\cr & \phantom{\tilde{\cal A}^{a+b+c}_{W\psi}  = }
+ 8\, D_{W\N\psi} \,m_{\psi} m_\N ( I_4^{[b]} + I_4^{[c]} ) \,, \cr
& \tilde{\cal A}^{a+b+c}_{\phi\psi}  =
c_\psi e_\psi^2 S_{\phi\N\psi} \left[ 2 m_{\psi}^2  \, I_4^{[b]} +
m_{\psi}^2 \, I_4^{[c]} + I_3^{[2]}(m_\phi, m_\psi) \right]
+ 2 c_\psi e_\psi^2 D_{\phi\N\psi} m_{\psi} m_\N \, I_4^{[b]}\,, \cr
& \tilde{\cal A}^{d}_{\phi\psi} = 
 \,c_\psi e_\psi^2\, g^P_{\phi\N\N}\,  
g^P_{\phi\psi\psi}\, {m_\psi \over m_\N}\, 
{4 m_\N^2 \over m_\phi^2-4m_\N^2} \, 
I_3^{[1]}(m_\psi)\,,\cr
& \tilde{\cal A}^e_{Z\psi} =  c_\psi e_\psi^2 \, g^A_{Z\N\N}
 \,g^A_{Z\psi\psi}\, {m_\psi^2 \over m_\N^2 } \,
   {4 m_\N^2 \over 4m_\N^2-m_Z^2}\, 
I_3^{[1]}(m_\psi)\,, \cr}
\equn\label{Result}
$$
where
$e_\psi$ is the electric charge of particle $\psi$ in units of the positron
charge, $c_\psi$ is a color factor (3 for quarks; 1 for leptons and charginos),
$$
\matrix{
S_{Wjk} = | g^V_{ijk} |^2 + | g^A_{ijk} |^2 \,, &
D_{Wjk} = | g^V_{ijk} |^2 - | g^A_{ijk} |^2 \,,
\cr\noalign{\vskip10pt}
S_{\phi jk} = | g^S_{ijk} |^2 + | g^P_{ijk} |^2 \,, &
D_{\phi jk} = | g^S_{ijk} |^2 - | g^P_{ijk} |^2 \,.
}
\equn
$$

The labels for amplitudes follow the labels appearing in
fig.~\ref{Diagrams}.  For example, ${\cal A}_{W\psi}^{a+b+c}$ is the gauge
invariant combination of diagrams appearing in
fig.~\ref{Diagrams}(a), (b) and (c) which have a $W$ boson in the
loop.  Similarly, ${\cal A}_{\phi \psi}^{a+b+c}$ is the same combination of
diagrams, but with the $W$ in the loop replaced by a scalar. 

For the triangle diagrams, $ \tilde{\cal A}^d_{\phi W}$ and $ \tilde{\cal
A}^d_{\phi \phi}$ do not contribute because of the special kinematics (see 
discussion in sect.~\ref{sect:kinematics}). $\tilde{\cal A}^e_{ZW}$,
$\tilde{\cal A}^e_{Z\phi}$ and the part of $\tilde{\cal A}^e_{Z\psi}$
with the vector coupling between $Z$ and the fermion loop vanish by Furry's
theorem. The term ${\cal A}^e_{Z\psi}$ requires special attention because of
the anomaly, which arises when the $Z$ coupling with the fermion loop
is axial. We have dropped the anomalous contributions in eqs.~(\ref{Result}) 
since they add to zero when summing over all fermions in the loop. The
remaining contribution of this diagram is non-zero due to the large mass
splitting between the two quarks of the third generation and between
the charginos.

Individual Feynman diagrams are, of course, not gauge-invariant.
However, the combinations appearing in eq.~(\ref{Result}) are gauge
invariant.  We have verified that when the photon polarizations are
taken to be longitudinal these combinations vanish as required by
gauge invariance.  This provides a strong check on our calculation,
including relative factors between diagrams.

Substituting the at rest kinematics (\ref{SpecialKinematics}) into
the general integral identity given in eq.~(18) of
ref.~\cite{IntegralsShort}, we can reduce the box integrals appearing in 
$\tilde{\cal A}^{a+b+c}_{W\psi}$ to triangle integrals: 
$$
\eqalign{
I_4^{[a]} & = {I_3^{[2]}(m_W, m_\psi)-I_3^{[1]}(m_W) \over m_\N^2+m_{\psi}^2-
            m_W^2}, \cr 
I_4^{[b]} & = {I_3^{[2]}(m_\psi, m_W)-I_3^{[1]}(m_\psi) \over m_\N^2+m_W^2-
            m_\psi^2}, \cr
I_4^{[c]} & = {I_3^{[2]}(m_\psi, m_W)-I_3^{[2]}(m_W, m_\psi) \over m_{\psi}^2-
            m_W^2}. \cr
}
\equn\label{AbbrIntegrals}
$$
The integral functions in $\tilde{\cal A}^{a+b+c}_{\phi \psi}$ are obtained
from eqs.~(\ref{AbbrIntegrals}) by replacing $m_W$ with $m_\phi$.

We use the standard methods of ref.~\cite{tHV} to express the
triangle integrals in terms of logarithms and dilogarithms
\cite{Dilogs},
$$
\eqalign{
I_3^{[1]}(m) & = \left\{ \eqalign{
{1\over 8 m_\N^2} \Bigl[ \log^2 \Bigl( {1+x\over1-x} \Bigr)
- \pi^2 - 2 i \pi \log \Bigl( {1+x \over 1-x} \Bigr) \Bigr]\,,
 \hskip 1.5 cm  & m \leq m_\N;\cr
- {1 \over 2 m_\N^2} \biggl( \arctan {1 \over \sqrt{m^2/m_\N^2 -1}} \biggr)^2 
\,,
\hskip 3.5 cm & m > m_\N; \cr } \right. \cr
I_3^{[2]}(m_1, m_2) & = 
{1\over 2 m_\N^2} \Bigl[\Li_2\Bigl({m_1^2-m_\N^2-m_2^2-\sqrt{\Delta_1} \over
2 m_1^2} \Bigr)
+\Li_2\Bigl({m_1^2-m_\N^2-m_2^2+\sqrt{\Delta_1} \over 2 m_1^2} \Bigr) \cr
& \hskip 1 cm
 -\Li_2\Bigl({m_1^2+m_\N^2-m_2^2-\sqrt{\Delta_2} \over 2 m_1^2}\Bigr)
-\Li_2\Bigl({m_1^2+m_\N^2-m_2^2+\sqrt{\Delta_2} \over 2 m_1^2} \Bigr) 
\Bigr], \cr
}
\equn\label{Triangles}
$$
where 
$$
\eqalign{
x & = \sqrt{1 - m^2 /m_\N^2} , \cr
\Delta_1 & = (m_1^2+m_\N^2-m_2^2)^2+ 4\, m_\N^2 m_2^2 , \cr
\Delta_2 & = (m_1^2-m_\N^2-m_2^2)^2- 4\, m_\N^2 m_2^2 . \cr }
\equn
$$
One convenient way to obtain the values of the integrals is from
Oldenborgh's program FF~\cite{Oldenborgh}.  In the notation used by FF
the triangle integrals are
$$
\eqalign{
& I_3^{[1]}(m) = I_3(m^2, m^2, m^2; 4 m_\N^2, 0, 0)\,, \cr
& I_3^{[2]}(m_1, m_2) = I_3(m_2^2, m_1^2, m_1^2; -m_\N^2, 0, m_\N^2) \,. \cr}
\equn
$$

We have compared the results in \eqn{Result} to those of BU~\cite{BU}
and we find complete agreement.  We also find disagreement between our
results and those of ref.~\cite{JK}. After using
eq.~(\ref{AbbrIntegrals}) to express the amplitudes in terms of
triangle integrals, the comparison is simple. For example, for the
integrals of class 1 and 2 of BU the translation between the real
parts of our integrals and those of BU is, $\Re I_3^{[1]}(m_\psi)
\leftrightarrow I_1(a,b)/4 m_\N^2$, $I_3^{[2]}(m_\psi, m_\phi)
\leftrightarrow - I_3(a,b)/2 m_\N^2$ and $I_3^{[2]}(m_\phi, m_\psi)
\leftrightarrow - I_2(a,b)/2 m_\N^2$.  (Notice that $I_3(a/b,1/b) =
I_2(a,b)$.) Similar formulas hold for the class 3 and 4 integrals.
The notation for the coupling constants is related by $g^L P_L + g^R
P_R = g^V + g^A \gamma_5$ (for vector bosons; $g^S + g^P \gamma_5$ for
scalar bosons) with $P_L = (1-\gamma_5)/2$ and $P_R =
(1+\gamma_5)/2$. (Note that BU use unitary gauge for the $Z$ in the
diagrams in fig.~\ref{Diagrams}(e), while we use Feynman-`t Hooft
gauge.)

\section{Discussion}

\subsection{Large mass behavior of amplitudes}

An interesting feature of our result is that in the limit when the
neutralino is a pure higgsino with a mass much greater than $m_W$ the
cross-section does not fall off when $m_\N$ is increased, but remains
constant. This behavior, which was numerically observed by BU, is
caused by the diagrams in \fig{Diagrams}(b). In the pure higgsino limit, $\N$
is almost degenerate with the lightest chargino and the
denominator of $I_4^{[b]}$ in eq.~(\ref{AbbrIntegrals}) becomes small,
so this contribution dominates. This allows us to obtain an analytic
expression in the limit $m_\N \simeq m_\ch \gg m_W$.  Using
eqs.~(\ref{AbbrIntegrals}),~(\ref{Triangles}), and the expansion
$$
\Li_2(1+i \eps)+\Li_2(1-i \eps) \simeq {\pi^2 \over 3} - \eps \pi,
\equn
$$
the asymptotic behavior of $I_4^{[b]}$ is
$$
I_4^{[b]} \simeq {\pi \over 2} {1 \over m_\N^3 m_W}.
\equn
$$
This yields the asymptotic behavior of the cross section, 
$$
\sigma v \simeq {\alpha^4 \pi \over 4 m_W^2 \sin^4\theta_W}   
\approx  10^{-28} \rm{cm}^3/\rm{s}.
\equn
$$
where we have used $S_{W \tilde{\chi} \tilde{\chi}^+} = D_{W
\tilde{\chi} \tilde{\chi}^+} = g^2/4 = \pi \alpha/\sin^2\theta_W$
and set the other couplings to zero which is appropriate for pure higgsinos.

\subsection{Annihilation to gluons}

We may also use the above results to obtain the cross section for
neutralino annihilation into two gluons.  The diagrams for this
process are a subset of the ones appearing in the photon
calculation. (The diagrams containing three-gluon vertices vanish.)
This allows one to obtain the result for gluon annihilation directly
from the results for photon annihilation.  For the gluon case, the
amplitude is given by
$$
\tilde{\cal A} = 
  \sum_{\matrix{\scriptstyle \phi=\tilde{q} \cr 
                 \scriptstyle \psi=q}}
\tilde{\cal A}^{a+b+c}_{\phi\psi} +
 \sum_{\matrix{\scriptstyle \phi=H^0_3,G^0 \cr 
               \scriptstyle \psi=q}}
\tilde{\cal A}^{d}_{\phi\psi} + 
\sum_{\psi=q} \tilde{\cal A}^{e}_{Z\psi} \, , 
\equn
$$
where we set $e_\psi=c_\psi=1$ in the individual amplitudes and sum
over quarks and squarks.  The color sum \cite{Bergstrom} amounts to
simply changing $\alpha^2$ to $2 \alpha_s^2$ in (\ref{CrossSection}),
so our result agrees with BU.  This confirms a small but significant
error in previous calculations \cite{Gluons,JKG}, which generally led to
an over-estimation of antiproton fluxes.\footnote{M. Kamionkowski has
reported to us that the error in the text of ref.~\cite{Gluons} did not
propagate to the numerical analysis of that paper.}

\section{Summary and Conclusions}

In this paper we presented a complete one-loop calculation for
neutralino annihilation into two photons or two gluons within the
Minimal Supersymmetric Standard Model.  In performing the calculation
we assumed that initial-state neutralinos are non-relativistic, which
is the relevant case for galactic dark matter.  The techniques we used
are, however, not limited to this special kinematics.

Our results confirm the very recent calculation of Bergstr{\"o}m and
Ullio \cite{BU} which disagrees with previously published ones, 
especially in the case where the neutralino is a heavy pure higgsino.

\vskip .8 cm 

We thank Graciela Gelmini for encouragement and a number of
stimulating discussions. This work was supported in part by the DOE
under contract DE-FG03-91ER40662 and by the Alfred P. Sloan Foundation
under grant BR-3222.  P.G. thanks Piero Ullio and Lars Bergst\"om for
useful discussions and David Cline, Graciela Gelmini and Roberto
Peccei for the extended visit at UCLA which made this work possible.

\end{document}